# TREND-BASED NETWORKING DRIVEN BY BIG DATA TELEMETRY FOR SDN AND TRADITIONAL NETWORKS

Ankur Jain[1], Arohi Gupta[2], Ashutosh Gupta[3], Dewang Gedia[4], Leidy Pérez[5], Levi Perigo[6], Rahil Gandotra[7] and Sanjay Murthy[8]

Interdisciplinary Telecom Program, University of Colorado Boulder, USA

## ABSTRACT

*Organizations face a challenge of accurately analyzing network data and providing automated action based on the observed trend. This trend-based analytics is beneficial to minimize the downtime and improve the performance of the network services, but organizations use different network management tools to understand and visualize the network traffic with limited abilities to dynamically optimize the network. This research focuses on the development of an intelligent system that leverages big data telemetry analysis in Platform for Network Data Analytics (PNDA) to enable comprehensive trend-based networking decisions. The results include a graphical user interface (GUI) done via a web application for effortless management of all subsystems, and the system and application developed in this research demonstrate the true potential for a scalable system capable of effectively benchmarking the network to set the expected behavior for comparison and trend analysis. Moreover, this research provides a proof of concept of how trend analysis results are actioned in both a traditional network and a software-defined network (SDN) to achieve dynamic, automated load balancing.*

## 1. INTRODUCTION

In the early-to-mid 1990s the Internet's growth changed the rules and due process in the daily lives of users all around the globe. What started as email for scientists and early file transfer systems, is now what moves every fiber of society regardless of culture, expertise or even direct exposure to connectivity [1]. This vast global adoption led to a shift in paradigm from manageable quantities of network data to a massive volume of traffic. Automation and orchestration applications such as virtual load balancers have become critical to cloud and data center control systems [2, 13, 15].

According to Cisco's figures for global IP growth for mobile and fixed networks, presented in their Visual Networking Index (VNI) Forecast report [3], the number of global Internet users will reach 4.6 billion by 2021 with 27.1 billion network devices, generating over 3.3 zettabytes
of IP traffic per year. Therefore, the Internet has entered big data telemetry territory, kicking off, as Cisco coined it first, The Zettabyte Era [4].

The main challenges that network managers are facing are scalability, analytics and actioning of results. Numerous Network Management System (NMS) tools have been developed with the goal of making data reveal the status of large networks. However, there is little automated action on the analytics results, which often include metrics on observable trends in network operations [5, 16].

The solution presented in this paper is an intelligent system that leverages the scalability and analytics capabilities of the Platform for Network Data Analytics (PNDA), to then enable automated networking decisions to be executed in the underlying network topologies.

The remainder of the paper is organized as follows: Section II provides a review of the existing body of knowledge, state of the art applications, and how our scheme extends it. Sections III and IV describe the methodology and results of our experiment respectively. Section V concludes our research and addresses scope for future enhancements.

## 2. RELATED WORK

When it comes to network analytics, there have been related efforts around providing a network manager with a global view of the performance of their network. The common output is a visual representation of data via dashboards and an alarm functionality for unexpected changes in performance. The end goal being passive monitoring that serves as an aid to network operations. However, the demand for functionality besides monitoring is overwhelming [6, 18, 19, 20, 21]. Enterprise Management Associates (EMA) did a research study titled "Network Management Megatrends 2016: Managing Networks in the Age of the Internet of Things, Hybrid Clouds and Advanced Network Analytics" [7]. Their premise was in this big data era, enterprise network managers are starting to have "higher aspirations" for the applications of the telemetry data collected from their networks. The results showed that, at the time of the study, advanced analytics tools have a deep impact in not only network operations, but in business applications. Network security monitoring resulted in the main advantage sought after with 38% of their data set, closely followed by network optimization as the second most popular use case with 32% of support. Their work not only backs up the multiple use cases for the research in this study, but also the increasing demand in network optimization driven by big data, specifically, big data telemetry. Big data telemetry defined as the robust collection of enormous quantities of network data and its aggregation onto one centralized big data analytics platform for analysis, which is the key decision driver of the architecture designed for the system presented in this paper. Furthermore, there are previous efforts aimed at providing a solution for a network topology that includes both traditional and software-defined networking (SDN) topologies [13, 15]. Additionally, network engineering practices face a major challenge of network optimization in this hybrid network scenario [8, 16].

One of these initiatives was brought forward by the Polytechnic School of Engineering from New York University. Their emphasis was to develop a congestion-aware single link failure recovery system for hybrid SDN [9], in order to achieve fast failover recovery and load balancing post recovery. The way they achieved this was by developing a heuristic algorithm that utilizes SDN devices in the recovery path that have a global view of the network to formulate recovery routes. The key difference from their work to the solution presented in this research is that they solved for a hybrid network, one with both traditional and SDN devices interfaced together, versus having a mixed solution by having the two network topologies be independent of each other, like the one this research focuses on.

In terms of automated load balancing, a team from the Ryazan State Radio Engineering University worked on developing an improved model of multipath adaptive routing in computer networks with a focus on load balancing capabilities [10]. Their approach on load balancing was heavily focused on the indicator of jitter optimization to select an optimal route. Their work resulted in an algorithm capable of acting as a baseline for new routing protocols, in contrast with the research for this study which presents a holistic user-oriented solution by integrating multiple subsystems to achieve a trend-based load balancing platform equipped with big data analytics and a Graphical User Interface (GUI) for ease of management.

In the article "A Dynamic Bandwidth Allocation Algorithm in Mobile Networks with Big Data of Users and Networks" [11] the authors utilize big data telemetry collected from the network to "cluster users by analyzing their closeness, both geographical and social, such that users in the same cluster share a wireless channel for downloading contents from the base station." Effectively showing how leveraging big data for dynamic resource allocation optimizes network management practices. However, the work done in this research differentiates itself from their research by having the core of the decision-making capabilities designed around trend analysis over time, creating an intelligent predictive system, versus real time resource allocation.

This research aims to answer the research question: **"Can big data collection, indexing, filtering, aggregation, benchmarking, and trend based analysis and actioning be service-chained to collectively provide a centralized big data analytics platform for telemetry data ingestion and analysis?"** This research question was strategically divided into sub-problems to answer the primary research question:

　　a. Can SNMP and OpenFlow be used to fetch network-wide telemetry data from traditional and software-defined networks respectively and store this information in a centralized database?

　　b. Can Logstash be used to filter and sort the data metrics to serve as input to PNDA's data streaming interface, Kafka? If so, can Jupyter Notebook streamline the data from Kafka interface to OpenTSDB through PNDA's network analysis tools?

　　c. Can statistical benchmarking be performed on the stored telemetry data by real-time monitoring to identify trends and take appropriate action?

The novel contribution of this paper is to design and implement a network architecture and an application that can be used together in a hybrid network to perform trend-based networking through analytics and proactive network monitoring. Furthermore, this research developed a self-sufficient system capable of making intelligent decisions through leveraging a powerful big data network analytics tool to load balance traffic in both SDN and traditional networks based on observable trends in link utilization. This research is beneficial to organizations because not only does it analyze network traffic, but also dynamically takes a corrective action for the identified network trend; thus, achieving trend-based, automated network optimization for better overall network performance.

## 3. RESEARCH OVERVIEW

The design solution for this system is developed for scalability, analytics, and actioning of results. All subsystems are carefully integrated to ensure telemetry collection is able to scale to massive data sets, given all hardware specifications are adjusted with the increase in traffic expectancy. The core component driving the design of all subsystems was PNDA, acting as the centralized big data analytics platform for telemetry data ingestion and analysis. See Figure 1.

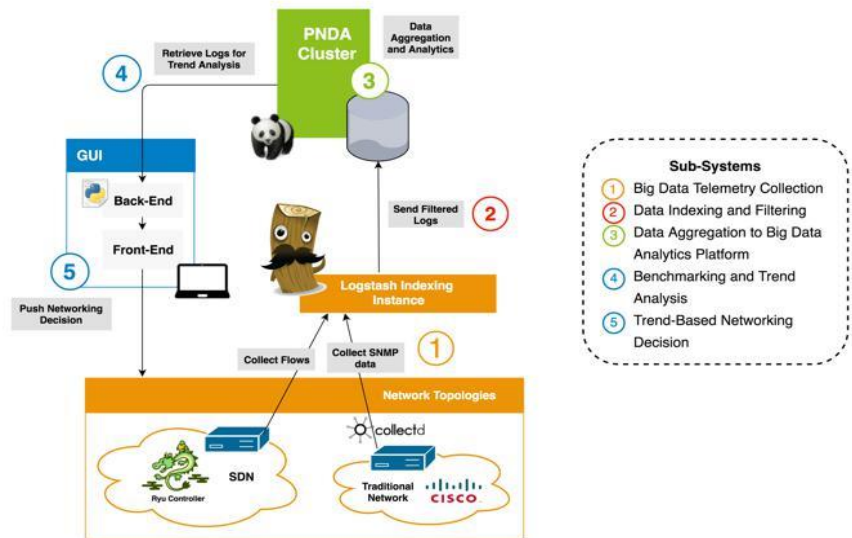

Fig. 1.  Test Architecture

### A. Can SNMP and OpenFlow be used to fetch network-wide telemetry data from traditional and software-defined networks respectively to store it in a centralized database?

The first step in the system is to perform the telemetry data collection from the underlying network topologies. This was achieved by designing a network configuration that meets the design requirements of SNMP data collection from traditional networks and switch statistics retrieval from SDN. SNMP data is collected from the traditional network using a systems statistics collection daemon (CollectD) to set frequency of data collection and tag traffic with the host's source IP. The traditional network topology can be seen in Figure 2, where SNMP agents have been set up in each device.

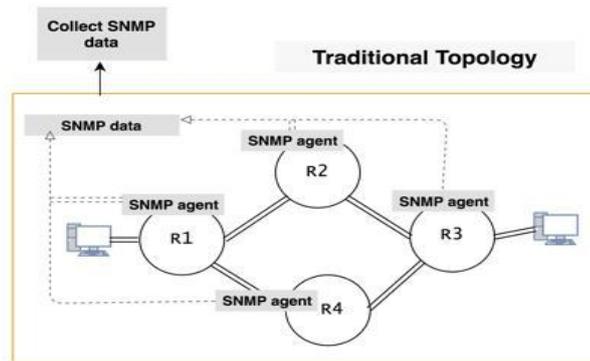

Fig. 2. Traditional Network Topology

The SDN topology is shown in Figure 3. To collect switch statistics from it, it is necessary to use a REST API and the OpenFlow protocol to fetch counters from each SDN switch via network flows. To meet this requirement, the selection of the SDN controller was based on the best fit REST API for the purpose, resulting in the implementation of the Ryu SDN controller. The resulting network architecture design enables a scalable telemetry data collection subsystem for both SDN and traditional networks.

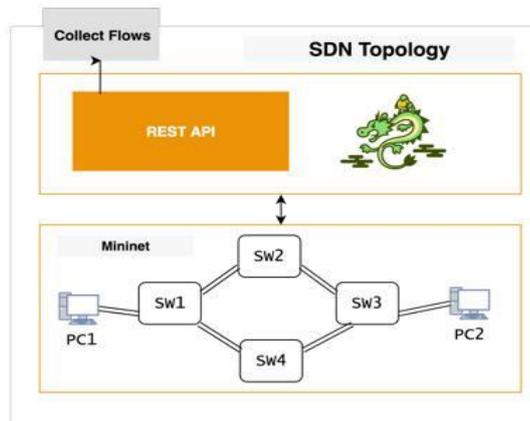

Fig. 3. SDN Topology

***B. Can Logstash be used to filter and sort the data metrics to serve as input to PNDA's data streaming interface, Kafka? If so, can Jupyter Notebook streamline the data from Kafka interface to OpenTSDB through PNDA's network analysis tools?***

The second step the system follows is to filter all telemetry data and prepare it for the big data analytics tool, PNDA. The design requirement was to have all raw data, collected from both networks, be filtered and formatted for seamless ingestion into PNDA. The tool selected for this development was Logstash, a data processing pipeline used to ingest data from different sources into a common platform with the required format [12].

For the traditional network, this is done by taking the data coming from the network via a CollectD configuration file written specifically for this project, and sending it to Logstash's input plugin. The SDN is capable of sending data directly into the same plugin without being filtered through CollectD.

Logstash was configured to run the data through its filter plugin and sort through it to get only specific metrics such as outPkts and outOctets. This data set is then sent to the output plugin, which serializes and encodes it using a codec named 'avro-codec', a requirement to ensure compatibility with the following subsystem. This sequence allows the system to send to PNDA's data streaming interface, Kafka, only valuable and properly formatted data, which in turn streamlines the process and reduces the workload on the server [17].

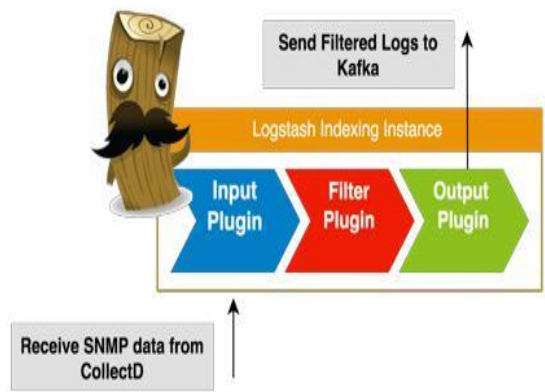

Fig. 4. Logstash Indexing Instance

The third step built into the system's workflow is the subsystem to aggregate all data coming from Logstash to PNDA. At this point, all data collected from both SDN and the traditional network is ready for PNDA's ingestion. The main design requirement for this stage is to get data from Kafka, the data streaming interface in charge of parsing through the data in real time, to PNDA's own Time Series Data Base, OpenTSDB. To do this, the main development was to implement an original Jupyter notebook.

A Jupyter notebook is the format of python algorithm Kafka is capable of integrating with. It contains all commands on what to do with input, data incoming from Logstash, and where to store the output. This stage is coded to go from Kafka, through PNDA's network analysis tools, to finally store results in OpenTSDB and make them available via Impala, a standard SQL interface used to query data from PNDA. See Figure 5.

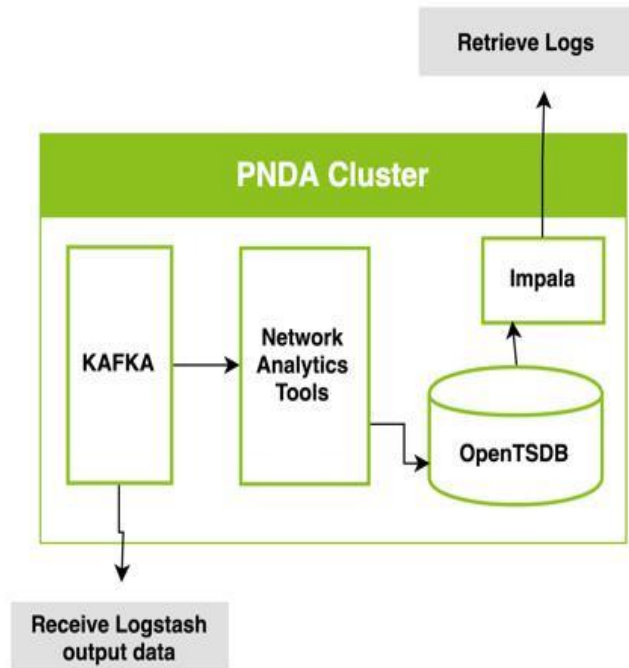

Fig. 5. PNDA Cluster

This results in data being ran through PNDA's analytics engine and getting ready to be utilized for benchmarking and trend analysis in the following subsystems.

*C. Can statistical benchmarking be performed on the stored telemetry data by real-time monitoring to identify the trend and take appropriate action?*

The fourth step for the integrated system to go through consists of leveraging PNDA's analytics to set a benchmark for both network configurations and to determine what behavior in the active topologies constitutes a trend that needs action in time versus changes that are one time occurrences. The development required for this sub-system was a GUI, see Figure 6, with two main components: a Python back-end architecture and a bootstrap front-end development. The front-end design, see Figure 7, is a comprehensive user-centric web application designed to address network engineers' needs of centralized resources to manage a system of this magnitude. To avoid duplicating efforts, it also has a redirect tab to PNDA's own dashboard for analytics, see Figure 8.

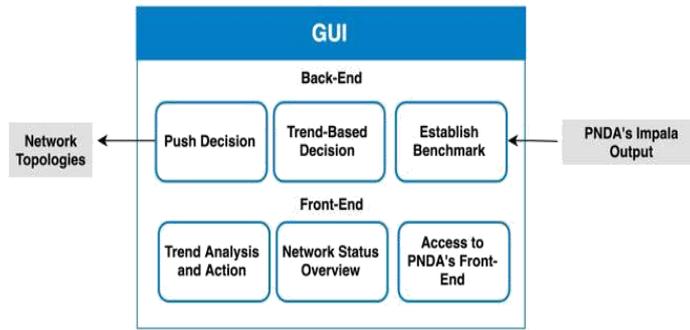

Fig. 6. Graphical User Interface (GUI) block diagram

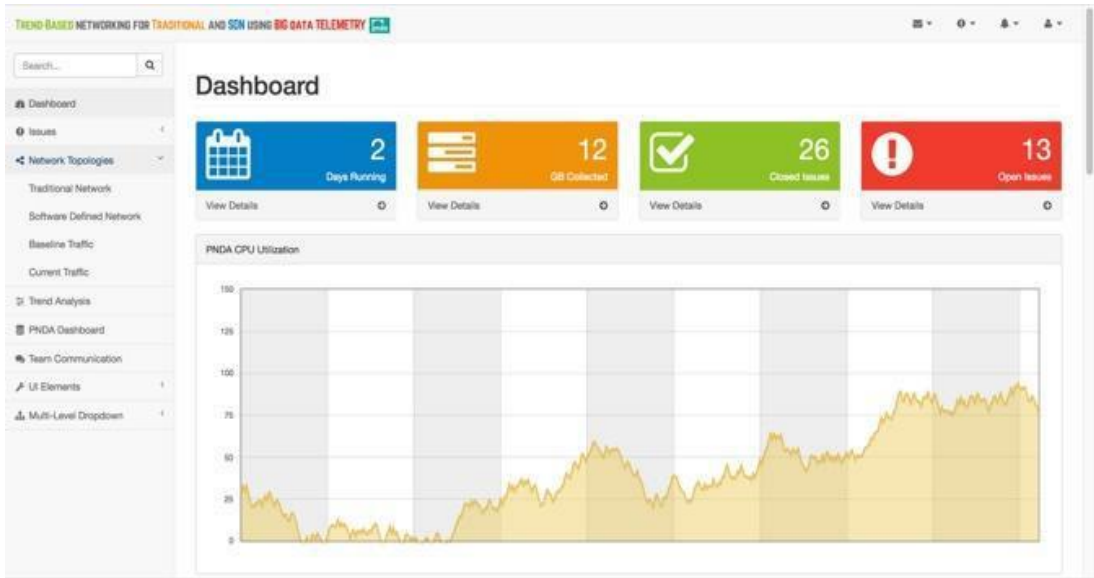

Fig. 7. GUI Front-End

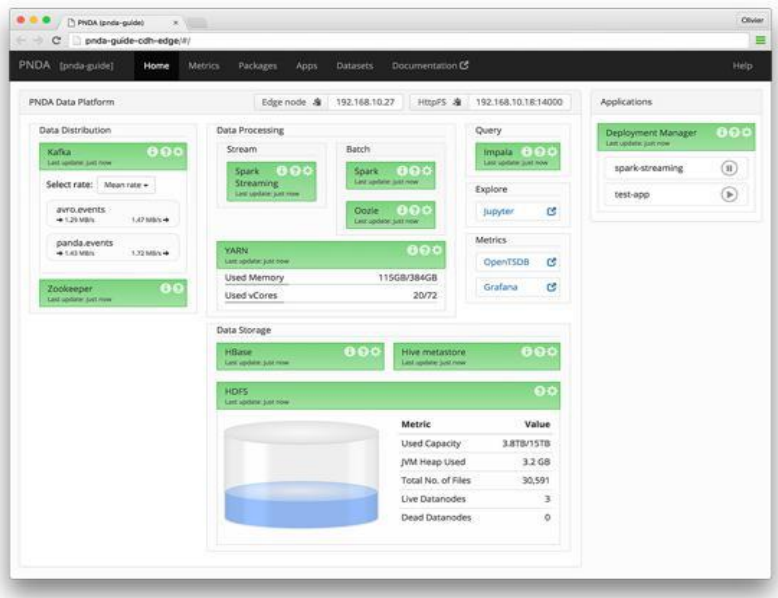

Fig. 8. PNDA's Dashboard

The back-end scheme is the engine that integrates all subsystems up to this step by performing two global tasks, benchmarking and trend analysis. Benchmarking runs all commands to execute subsystems one through three to then consume data analytics coming from PNDA and execute the algorithm's section developed to evaluate network behavior without intervention. This unit of the code is set to query all telemetry data for an established time period and frequency, to then run a statistical distribution of data and get the mean value of link utilization for the main output ports in the system. It will also set a threshold of interface bandwidth to separate "normal traffic," from "high traffic," which will also be used as a trend indicator. This process will repeat to reset the benchmark with a user determined frequency. See Figure 9.

For trend analysis, the code was developed to identify link utilization trends that will be utilized for load balancing. It takes the statistical values and threshold established during benchmarking to find the standard deviation of new incoming traffic, all while monitoring the interface bandwidth threshold. Once traffic crosses the threshold and deviates significantly from mean value of traffic, the code will raise a flag to mark the trend, until traffic lowers back to below the threshold or within regular values, and the change that needs to be addressed is established in a timeframe. This process effectively enables the back-end to action a specific trend and make the right decision to load balance the system during the right period of time.

Once data is collected, filtered, ran through big data analytics and submitted for benchmarking to identify the trend, the system is ready for the call to action on the latest trend spotted by the back-end, all while providing visibility in the front-end. The call to action of a trend-based networking decision is executed by pushing the appropriate data structure to the correct devices in each topology. For the traditional network, this is done via an open-source library that enables SSH to routers, Netmiko. Which will send the appropriate route-map CLI commands to the key affected router. The route-map contains the CLI commands necessary for traditional routers to associate a lower local preference for the specific affected subnet and a higher local preference for the new path, which determines where traffic is routed during the timeframe needed. For the SDN topology, the same concept is used, with the difference of flows entries being pushed to the SDN instead of route-maps.
.

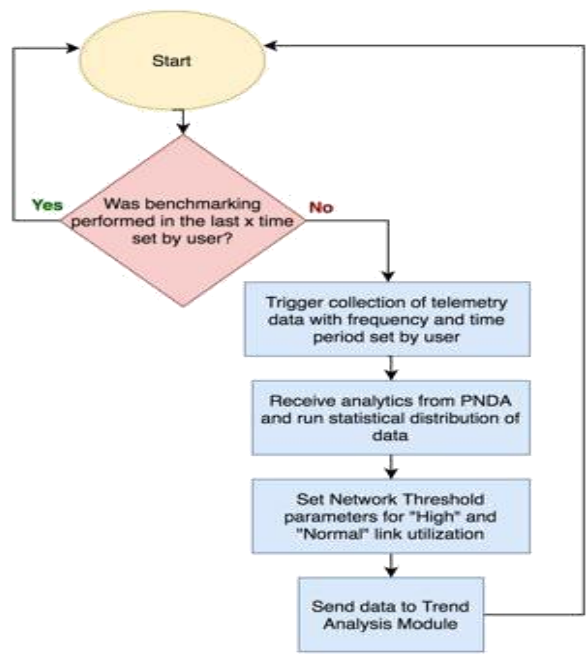

Fig. 9. Benchmarking Code Flow

OpenFlow flow entries will determine the change in path for traffic during the timeframe established. Once a decision has been made and pushed to the network, the user can choose to have the decision be effective for a period of time and then revert it to reevaluate the status over time in their network. The logic followed by the code is detailed in Figure 10.

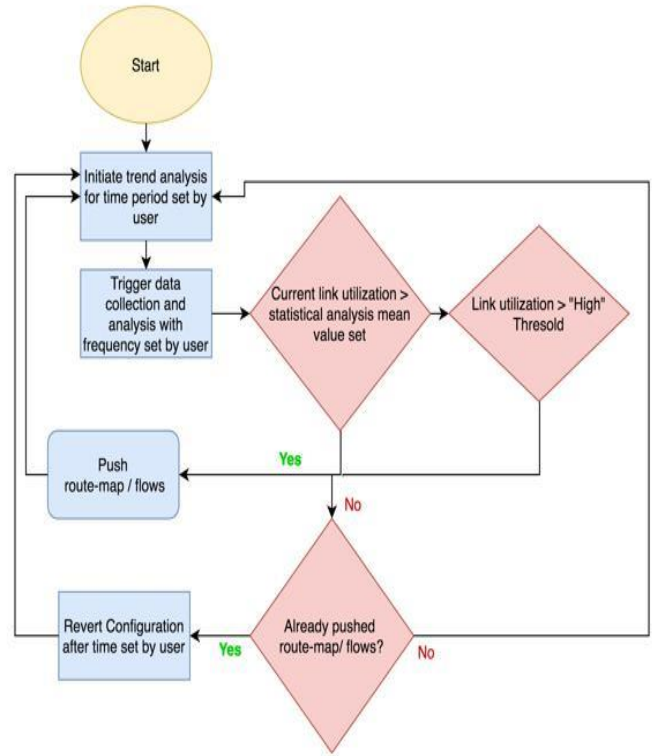

Fig. 10. Trend-Based Networking Decision Block Diagram

When all steps are performed in an integrated, seamless manner, the end result is a trend-based networking system that utilizes powerful tools, such as PNDA for big data telemetry analysis, to provide the user with the most powerful interpretation and usage of their network data. Not only does the user get all monitoring benefits from PNDA's built in functionality, but also gets a GUI on which to track trends and actions taken in the network to keep traffic running without congestion or packet loss, with minimal human intervention.

## 4. RESULTS AND ANALYSIS

### A. Physical Test Setup

Two servers were utilized to distribute all components of the system as seen in Figure 11. The hardware and software specifications are broken down in Table 1 and Table 2 respectively.

Multiple levels of virtualization and different types of servers were used to complete this physical setup. The Server 1 is a bare metal server running Ubuntu 16.01 as its Virtualization Layer 0. On this Ubuntu Server is where VirtualBox is set up, making Virtualization Layer 1. Then, VirtualBox has Ubuntu 14.04 installed as Virtualization Layer 2, which contains the PNDA VM and the GUI as its logical components. Server 2 is dedicated to network virtualized topologies. This bare metal server runs VMware Esxi 6.5 as its Virtualization Layer 0 and two Virtual Machines (VMs) running Ubuntu 16.01, Virtualization Layer 1, for each network design.

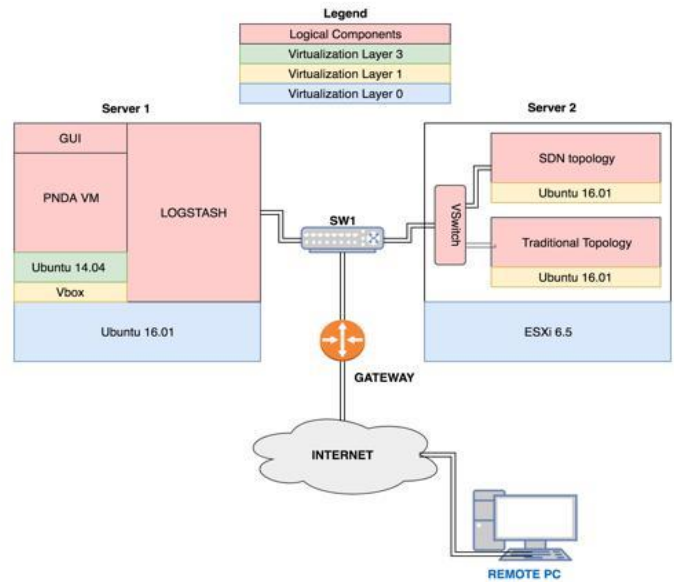

Fig. 11. Physical setup

Table I. Hardware Specifications

| Device | Hardware Specifications | | | |
|---|---|---|---|---|
| | *Make* | *Model* | *CPU* | *Memory* |
| Servers | Dell Inc | PowerEdge R430 | 12 CPUs x Intel(R) Xeon(R) CPU E5-2603 v3 @ 1.60GHz | 32 GB |

Table II. Software Specifications

| Component | Software Specifications |
|---|---|
| Server 1 | Version |
| Ubuntu | 16.01 |
| Ubuntu | 14.04 |
| VirtualBox | 4.0 |
| Logstash | 2.4.1 |
| Red-PNDA | 0.2.4 |
| Server 2 | Version |
| Ubuntu | 16.01 |
| GNS3 | 0.8.7 |
| Cisco 7200 | IOS: 15.2(4)S5 |
| Ubuntu | 16.01 |
| RYU | 4.23 |
| Mininet | 2.3.0d1 |

The first VM has the traditional network topology generated using GNS3. The second VM has the SDN components, in which the Ryu controller and Mininet are set up to establish the topology. These two VMs are connected to each other using a virtual switch, and further connected to a router which allows connectivity to the Internet using port forwarding.

## B. Proof of Concept

**Establish Benchmark**

In order to determine the mean value of link utilization of output ports, it was necessary to first perform telemetry data collection and analysis of the network in a period of 72 hours, in order to get the baseline behavior. This was achieved by implementing the sub-system integration from step 1 through step 4 of the workflow described in section III A, Figure 1.

First, traffic was generated using a network performance tool called iPerf 3. The frequency was set to every 1 hour, 24 samples per day, and 72 samples in total. The network topologies were set up as displayed in Figures 2 and 3.

The assumption made for traffic generation was that the highest activity hours are between business hours, 8 AM to 5 PM, with activity decreasing significantly outside of that time frame.

Then for step two, as traffic was generated, it was aggregated by CollectD to Logshtash's input plugin for the traditional network and directly from the RYU controller REST API. Logstash ran the data through its filter plug in to relay inPkts, outPkts, inOctets, outOctets, interface name and time stamps to PNDA's Kafka interface using the pnda-avro plugin. See Figure 12 for a sample of the output coming from Logstash's output plugin.

```
{
           "@timestamp" => "2018-04-07T22:05:24.218Z",
               "plugin" => "snmp",
        "collectd_type" => "if_cols",
        "type_instance" => "FastEthernet0_1",
             "inOctets" => 18050215,
               "inPkts" => 77062,
            "inDiscards" => 0,
            "outOctets" => 18102066,
              "outPkts" => 77063,
           "outDiscards" => 0,
             "@version" => "1",
                  "src" => "collectd",
              "host_ip" => "10.0.0.10",
              "rawdata" => "\"{ \\\"interface\\\" : \\\"FastEthernet0_1\\\", \\
\"inPkts\\\" : 77062,  \\\"outPkts\\\" : 77063, \\\"inOctets\\\" : 18050215
, \\\"outOctets\\\" : 18102066}\"",
            "timestamp" => 1523138724218
}
```

Fig. 12. Output from Logstash's Output Plugin

Now the system is at step three and traffic has arrived to PNDA. Once the Kafka interface detects the data input, a Jupyter query takes the telemetry data and sends it to the OpenTSDB database for storage. A snippet of data stored in OpenTSDB can be seen in Figure 13. This shows how once telemetry data gets to this stage, it is already separated by type. For this snippet, the display is a snapshot in time for outOctets.

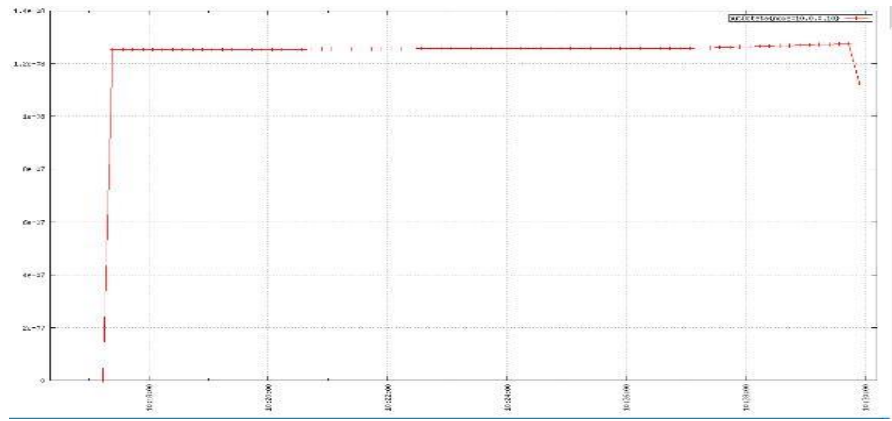

Fig. 13. Snippet of outPkts in OpenTSDB

Once telemetry data is analyzed and separated by type of input or output, comes the fourth step in the systems sequence, which consists of the back-end application retrieving data logs from Impala to set the benchmark for trend analysis. The results of the benchmarking in the back-end are shown in Figure 14.

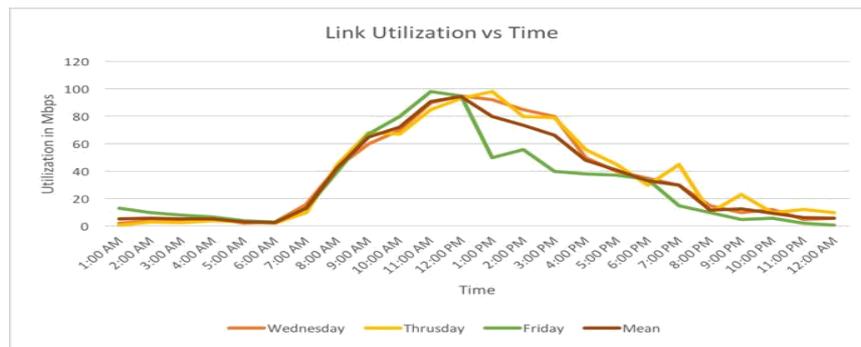

Fig. 14. Benchmarking performed in the back-end

From the graph, it can be observed how the data samples were collected for three (3) days with a frequency of once per hour. A mean value was calculated for every hour separately. This mean value is used as the benchmark for that particular hour and will be used in the trend based analysis in future. These results enable the second round of testing, which will entail taking this benchmark to identify trends and send automated corrective action back to the topologies, fulfilling the last step in the system's workflow.

## 5. CONCLUSIONS AND FUTURE WORK

In this paper, all steps to achieve the full integration of the trend-based networking system driven by big data telemetry for SDN and traditional networks have been described. It has been proven that benchmarking the network using all the subsystems built in the design solution is possible and effective. Moreover, the prospect of the tests conducted in this research can further be calibrated based on the desired use-case for automated load balancing and trend-based networking. The contributions of this research relies on the algorithms carefully developed to run subsystems at the top of their performance, while integrating seamlessly to create a fully integrated system. The user gets both the benefits of monitoring all the data PNDA is analyzing via PNDA's own dashboard, plus an automated trend-based load balancing system administered via the system' own GUI with little human intervention needed.

This design sets the foundation for the next generation of network management system tools. The number of applications that can be built on top of this work is only limited by what application the user envisions in their network. Some of the next steps and applications that could be implemented are:

A. With the benchmark set in the performed test along with the mean value and link utilization threshold, the next step is to collect data again over a span of seven days, this time picked tentatively as a generic number for testing purposes. The frequency of collection will be the same used in benchmarking, once an hour, to keep compatibility between data sets. Once the telemetry data set has gone through all the same subsystems that the benchmarking test used, it will be time to identify the trend by analyzing the standard deviation of the data sample for every hour against its counterpart in the benchmark data set, following the system logic detailed in Figure 10.

B. Expand the scope of network metrics used for trend-based decisions: link utilization was used as a proof of concept. This network can be scaled to include any network metric compatible with the data collection system, such as CPU utilization, bandwidth, and latency.

C. Machine Learning for Network Trends: leveraging PNDA's big data capabilities, it is possible to keep a comprehensive archive of all trends identified and have a machine learning algorithm running in the back end to start predicting network behavior before the trend even occurs.

D. Evolve from load balancing via prefix to load balancing via multipath: our current solution implements a prefix-centric load balancing approach. If this is taken one step further, multipath BGP can be implemented for further load balancing capabilities.

E. Connecting traditional and SDN: while this solution is oriented towards SDN and traditional networks operating independently, it can be developed to support both networks communicating enabling all hosts pertaining to both networks to achieve end-to-end connectivity.

**AUTHORS**

**Ankur Jain** is a graduate student from the University of Colorado Boulder with a major in Network Engineering. He currently works as an Associate Systems Engineer at Juniper Networks.

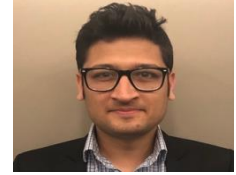

**Arohi Gupta** is a graduate student from the Interdisciplinary Telecom Program, University of Colorado Boulder. She received her bachelor's degree in Electronics and Telecommunications Engineering, and currently works as Software Engineer 3 at Juniper Networks.

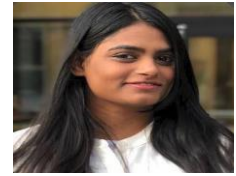

**Ashutosh Gupta** is a graduate student from the University of Colorado Boulder with a major in Network Engineering. He received his bachelor's degree in Electronics and Telecommunications Engineering, and currently works as SDN/NFV Software Engineer 3 at Juniper Networks.

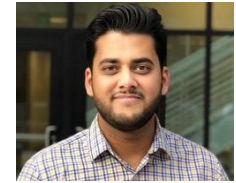

**Dewang Gedia** is a Ph.D. candidate at the Interdisciplinary Telecom Program, University of Colorado Boulder. He received his master's degree in Network Engineering, and has primary research focus in network functions virtualization and software-defined networks domain.

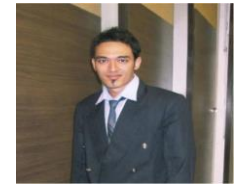

**Leidy Pérez** is a graduate student from the Interdisciplinary Telecom Program, University of Colorado Boulder. She received her bachelor's degree in Telecommunications Engineering, and currently works as Strategic Operations Manager at Zayo Group.

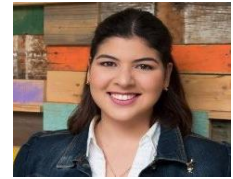

**Dr. Levi Perigo** is a Scholar in Residence and Professor of Network Engineering at the Interdisciplinary Telecom Program, University of Colorado Boulder. His interests are in a variety of internetworking technologies such as network automation, VoIP, IPv6, SDN/NFV, and next generation protocols. Currently, his research focuses on implementation and best practices for network automation, SDN, and NFV.

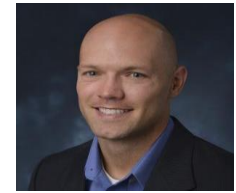

**Rahil Gandotra** is a Ph.D. student at the Interdisciplinary Telecom Program, University of Colorado Boulder. He received his bachelor's degree in Telecommunications Engineering, and has primary research interests in next-generation networking focusing on software-defined networking, network functions virtualization, and energy-efficient networking.

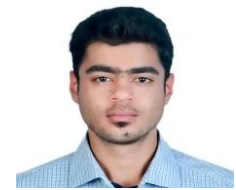

**Sanjay Murthy** is a graduate student from the Interdisciplinary Telecom Program, University of Colorado Boulder. He received his bachelor's degree in Electronics and Communications Engineering, and currently works as Production Network Engineer at Facebook.

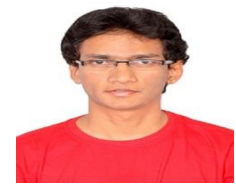